\documentclass[useAMS,usenatbib]{mnras}

\usepackage{graphicx}
\usepackage[switch,modulo]{lineno}
\usepackage{amsxtra}
\usepackage{amssymb}

\usepackage{times}
\usepackage{hyperref}

\usepackage{breakurl}
\title[Outburst of CTA~102 in 2012.]{Exceptional outburst of the blazar CTA~102 in 2012: The GASP-WEBT campaign and its extension.
\thanks{The radio-to-optical data collected by the GASP-WEBT collaboration are stored in the GASP-WEBT archive;
for questions regarding their availability, please contact the WEBT president, Massimo Villata ({\tt villata@oato.inaf.it}).}}

   \author[V.M.\,Larionov et al.]
	{\parbox{\textwidth}{V.M.\,Larionov,$^{1,2}$\thanks{e-mail {\tt v.larionov@spbu.ru}}
M.\,Villata,$^{3}$
C.M.\,Raiteri,$^{3}$
S.G.\,Jorstad,$^{1,4}$
A.P.\,Marscher,$^{4}$
I.\,Agudo,$^{5}$
P.S.\,Smith,$^{6}$
J.A.\,Acosta-Pulido,$^{7,8}$
M.J.\.Ar\'evalo,$^{7,8}$
A.A.\,Arkharov,$^{2}$
R.\,Bachev,$^{9}$
D.A.\,Blinov,$^{1,10,11}$
G.\,Borisov,$^{9,12}$
G.A.\,Borman,$^{13}$
V.\,Bozhilov,$^{14}$
A.\,Bueno,$^{7,8}$
M.I.\,Carnerero,$^{3,7,8}$
D.\,Carosati,$^{15,16}$
C.\,Casadio,$^{5}$
W.P.\,Chen,$^{17}$
D.P.\,Clemens,$^{4}$
A.\,Di Paola,$^{18}$
Sh.A.\,Ehgamberdiev,$^{19}$
J.L.\,G\'omez,$^{5}$
P.A\,Gonz\'alez-Morales,$^{7,8}$
A.\,Gri\~n\'on-Mar\'in,$^{7,8}$
T.S.\,Grishina,$^{1}$
V.A.\,Hagen-Thorn,$^{1}$
S.\,Ibryamov,$^{9}$
R.\,Itoh,$^{20}$
E.N.\,Kopatskaya,$^{1}$
E.\,Koptelova,$^{17}$
C.\,L\'azaro,$^{7,8}$
E.G.\,Larionova,$^{1}$
L.V.\,Larionova,$^{1}$
A.\,Manilla-Robles,$^{7,8}$
Y.\,Metodieva,$^{14}$
Yu.V.\,Milanova,$^{1}$
D.O.\,Mirzaqulov,$^{19}$
S.N.\,Molina,$^{5}$
D.A.\,Morozova,$^{1}$
S.V.\,Nazarov,$^{13}$
E.\,Ovcharov,$^{14}$
S.\,Peneva,$^{9}$
J.A.\,Ros,$^{21}$
A.C.\,Sadun,$^{22}$
S.S.\,Savchenko,$^{1}$
E.\,Semkov,$^{9}$
S.G.\,Sergeev,$^{13}$
A.\,Strigachev,$^{9}$
Yu.V.\,Troitskaya,$^{1}$
I.S.\,Troitsky,$^{1}$
}
\vspace{0.4cm}\\
\parbox{\textwidth}{
$^{1}$Astronomical Institute, St.-Petersburg State University, St.-Petersburg, Russia\\
$^{2}$Pulkovo Observatory, St.-Petersburg, Russia\\
$^{3}$INAF, Osservatorio Astrofisico di Torino, Italy\\
$^{4}$Institute for Astrophysical Research, Boston University, Boston, MA, USA\\
$^{5}$Instituto de Astrofis\'ica de Andaluc\'ia, CSIC, Granada, Spain\\
$^{6}$Steward Observatory, University of Arizona, Tucson, AZ85721, USA\\
$^{7}$Instituto de Astrofisica de Canarias (IAC), La Laguna, Tenerife, Spain\\
$^{8}$
Departamento de Astrofisica, Universidad de La Laguna, La Laguna, Tenerife, Spain
\\
$^{9}$Institute of Astronomy, Bulgarian Academy of Sciences, Sofia, Bulgaria\\
$^{10}$
Department of Physics and Institute for Plasma Physics, University of Crete, 71003, Heraklion, Greece
\\
$^{11}$
Foundation for Research and Technology - Hellas, IESL, Voutes, 7110 Heraklion, Greece
\\
$^{12}$Armagh Observatory, Northern Ireland, UK\\
$^{13}$Crimean Astrophysical Observatory, Russia\\
$^{14}$Dept. of Astronomy, Faculty of Physics, Sofia University, Bulgaria\\
$^{15}$EPT Observatories, Tijarafe, La Palma, Spain\\
$^{16}$INAF, TNG Fundacion Galileo Galilei, La Palma, Spain\\
$^{17}$Graduate Inst.\ of Astronomy, National Central Univ., Jhongli, Taiwan\\
$^{18}$INAF, Osservatorio Astronomico di Roma, Italy\\
$^{19}$
Maidanak Observatory of the Ulugh Beg Astronomical Institute, Uzbekistan\\
$^{20}$Department of Physical Sciences, Hiroshima University, Japan\\
$^{21}$Agrupaci\'o Astron\`omica de Sabadell, Spain\\
$^{22}$Department of Physics, University of Colorado, Denver, USA\\
						}}
\date{Accepted 2016 ???. Received ???; in original form 2016 ???}
\pubyear{2016}	
\begin{document}
\label{firstpage}
\pagerange{\pageref{firstpage}--\pageref{lastpage}} 
\maketitle

\clearpage
\begin{abstract}%
After several years of quiescence, the blazar CTA 102 underwent an
exceptional outburst in  2012 September--October. The flare was tracked from $\gamma$-ray to near-infrared frequencies, 
including \textit{Fermi} and \textit{Swift} data as well as photometric and polarimetric data from several observatories.
An intensive GASP-WEBT collaboration campaign in optical and NIR bands, with an addition of previously unpublished
archival data and extension through fall 2015, allows comparison of this outburst with the previous activity period
of this blazar in 2004--2005. We find remarkable similarity between the optical and $\gamma$-ray behaviour of CTA~102 
during the outburst, with a time lag between the two light curves of $\approx 1$ hour, indicative of co-spatiality of 
the optical and $\gamma$-ray emission regions. The relation between the $\gamma$-ray and optical fluxes is consistent with
the SSC mechanism, with a quadratic dependence of the SSC $\gamma$-ray flux on the synchrotron optical flux evident in the
post-outburst stage. However, the $\gamma$-ray/optical relationship is linear during the outburst; we attribute this
to changes in the Doppler factor. A strong harder-when-brighter spectral dependence is seen both the
in $\gamma$-ray and optical non-thermal emission. This hardening can be explained by
convexity of the UV--NIR spectrum that moves to higher frequencies owing to an increased
Doppler shift as the viewing angle decreases during the outburst stage. 
The overall pattern of Stokes parameter variations agrees with a model of a radiating blob or shock wave that moves
along a helical path down the jet. 
\end{abstract}

%
%
\section{Introduction}
\label{intro}
The blazar CTA~102 (4C +11.69, 2FGL J2232.4+1143, $z=1.037$) is a luminous, well-studied
quasar. Like other blazars, it is believed that its jet is oriented close to our line of sight, which causes
strong relativistic beaming of the jet's emission and violent variability at all wavelengths. CTA~102 was first
identified as a quasar by \citet{Sandage1965}
and belongs to the optically violently variable (OVV) \citep{Angel1980}, as
well as the high polarized quasar (HPQ), subclasses \citep{Moore1981}.

On long time-scales the blazar exhibits rather modest variability at optical
bands. Moderate-amplitude fluctuations around the average magnitude of $B=17\fm7$ over a 14 yr range (about 65 observations
between 1973 and 1987) were reported by \citet{Pica1988}. An
overall amplitude $\Delta R=0\fm88$ was observed by \citet{Villata2001} in
1994-1997. However, occasional sharp flares have also been observed in CTA~102.
Variations as high as $\Delta B=1\fm07$ in 2 days \citep{Pica1988} and $\Delta
V = 1\fm13$ in 3 days \citep{2000A&AS..143..357K} were observed in 1978 and 1996,
respectively. The previously reported historical maximum for the object,
$R \approx 14\fm5$, was reached on Oct. 4, 2004 during a short-term event
accompanied by prominent intra-night variability \citep{Osterman2009}. Between that episode
and 2012, only moderate variability has been seen in the light curve of this blazar (see Fig.~\ref{lc_total}).

CTA~102 was discovered to be a $\gamma$-ray emitter early in the
{\em Compton} Gamma Ray Observatory (CGRO; EGRET detector) mission at a level of $(2.4 \pm 0.5) \times 10^{-7} \mbox{ph }
\mbox{cm}^{-2} \mbox{s}^{-1}$ ($E>100$ MeV) \citep{Nolan1993}. It was also
detected in the 10--30 MeV energy range by the COMPTEL instrument of CGRO
\citep{Blom1995}. Since the blazar usually exists in a quiescent state, the
average $\gamma$-ray flux is rather low, $(2.9\pm0.2) \times 10^{-9} \mbox{ph }
\mbox{cm}^{-2} \mbox{s}^{-1}$ ($1<E<100$ GeV) according to the 2FGL catalog based on data from
the Large Area Telescope (LAT) of the {\em Fermi} Gamma-ray Space Telescope
\citep{Nolan2012}. Therefore,  accurate relative timing of flux variations in
$\gamma$-ray and optical bands is only possible during large outbursts. Similar events may serve as a
crucial test for models localizing the $\gamma$-ray emission in blazars  \citep[e.g.,][]{Marscher2010b}.
This type of cross-correlation analysis, performed for several other blazars, has recently shown that
$\gamma$-ray and optical flares are usually coincident \citep[e.g.,][]{Raiteri2012, Raiteri2013} and associated with the
passage of a new superluminal knot through the 43-GHz radio core \citep[e.g.,][]{Marscher2010, Agudo2011}.
\citet{Casadio2015} studied the evolution of the parsec-scale jet in CTA~102 with  ultra-high angular resolution through a
sequence of 80 total and polarized intensity
Very Long Baseline Array (VLBA) images at 43 GHz, covering the time span from June 2007 to June
2014. They have shown that a flare seen both in $\gamma$ and optical bands took place $\ga$12~pc from the black hole,
and suggested the synchrotron self-Compton (SSC) process as the source of the $\gamma$-ray emission.

In this paper we analyse the largest outburst of CTA~102 to date at optical and $\gamma$-ray bands \citep{Larionov2012}.
A preliminary analysis of our data collected through fall 2012 is reported in~\cite{Larionov2013b};
in the present paper we extend the data set up to the end of 2015. 
In \S~\ref{observ} we describe our observational data and their reduction; in \S~\ref{color_evolution} we analyse the
colour variability of CTA~102 and evolution of its spectral energy distribution (SED); \S~\ref{gam_opt} deals with
$\gamma$ -- optical correlations. Optical spectra are discussed in \S~\ref{spectra}. The polarimetric behaviour of this
blazar and a model describing its temporal evolution are presented in \S~\ref{discussion}. The final conclusions
are summarised in \S~\ref{conclusions}.
\section{Observations and data reduction}
\label{observ}

\subsection{Optical and Near-infrared Photometry}

The GASP-WEBT ~\citep[see e.g.,][]{2008A&A...481L..79V, 2009A&A...504L...9V} observations in 2008--2013 were performed in
$R$ band at the following observatories: Belogradchik, Calar Alto, Crimean Astrophysical, Lowell (Perkins telescope), Lulin,
Mount Maidanak, New Mexico Skies (iTelescopes), Roque de los Muchachos (Liverpool Telescope), Rozhen, Sabadell, Skinakas,
St.~Petersburg, Teide (IAC80), and Tijarafe. $BVI$ photometric data are from St.~Petersburg and Lowell observatories.
The $V$ and $R$-band light curves are complemented by data taken at Steward Observatory under a monitoring program in
support of the \textit{Fermi} mission. Near-infrared (NIR, $JHK$) data are from the Perkins Telescope,
AZT-24 (Campo Imperatore), and Teide (TCS). We also use $B$ and $R$ Mt.~Maidanak data during the 2004 outburst. After
the nominal end of the GASP campaign, we continued monitoring CTA~102 in optical--NIR bands (Crimean Astrophysical Observatory,
Lowell Observatory, St.~Petersburg University, Steward Observatory, Campo Imperatore observing station of Rome Observatory) in order to track the post-outburst behaviour.
We used photometric sequences in optical bands reported in \cite{1998A&AS..130..495R} and, in NIR bands, those given on
the AZT-24 web-page \footnote{\url{http://www.astro.spbu.ru/staff/vlar/NIRthumbs/cta102.html}}. 

We corrected the optical and NIR data for Galactic extinction using values reported in the NASA Extragalactic Database
(NED)\footnote{\url{http://ned.ipac.caltech.edu/}} for each filter \citep{Schlafly2011}.
The magnitude to flux transformations were calculated with coefficients determined by \cite{Mead1990}.

\begin{figure}
\begin{center}
   \includegraphics[width=\columnwidth,clip]{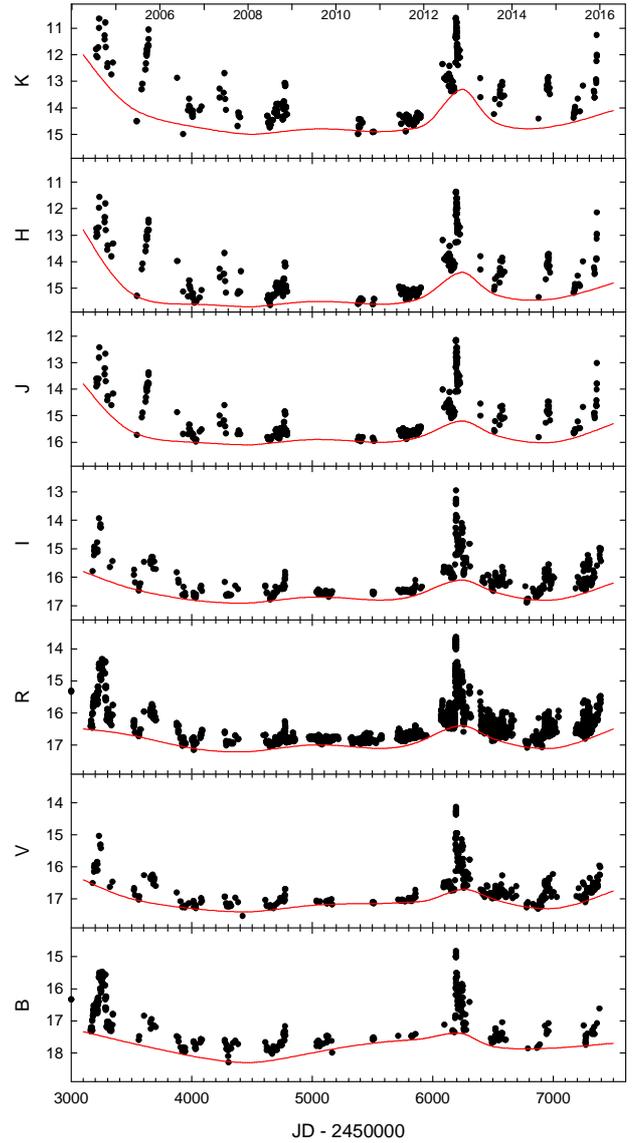}
    \caption{Optical and near-infrared light curves of CTA~102 over the time interval 2004--2015.
    Spline curves correspond to lower envelopes for variations in each colour band. Hereafter we denote TJD=JD-2450000.0.}
\label{lc_total}
\end{center} 
\end{figure}

The optical and near-IR light curves of CTA~102 during the 2004--2015 time interval
are shown in Fig.~\ref{lc_total}; spline curves correspond to lower envelopes of variations in each colour band.
We note that during both the 2004 and 2012 outbursts the amplitudes of long-term (marked with splines)  and  short-term
(individual data points) variations increase with wavelength, as is common in flat-spectrum radio quasars (FSRQs). 

\subsection{Optical Polarimetry}
\begin{figure}
\begin{center}
   \includegraphics[width=\columnwidth,clip]{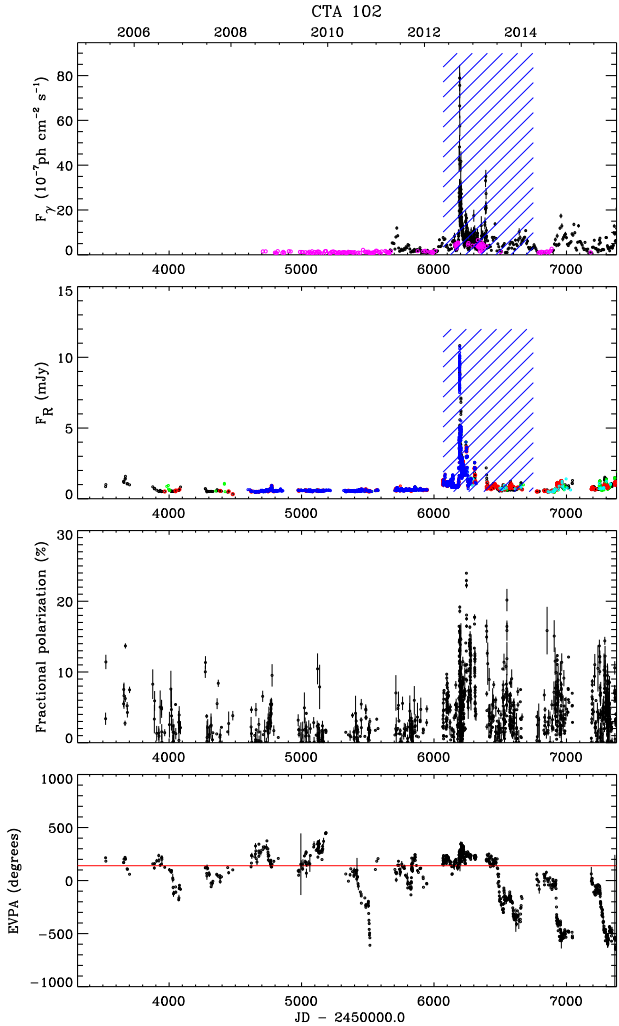}
    \caption{From top to bottom: $\gamma$-ray and optical flux evolution, optical fractional polarization, and
    position angle of polarization of CTA~102 over the time interval 2005--2015. Magenta points in the $\gamma$-ray
    light curve indicate upper limits; blue symbols in the optical panel denote GASP data. Shaded areas in two upper panels mark the outburst time interval, as discussed in \S~\ref{gam_opt}. The red line in the EVPA panel corresponds to the mean direction of mm-wave radio jet.}
\label{cta102lc1}
\end{center} 
\end{figure}

\begin{figure}
\begin{center}
   \includegraphics[width=\columnwidth,clip]{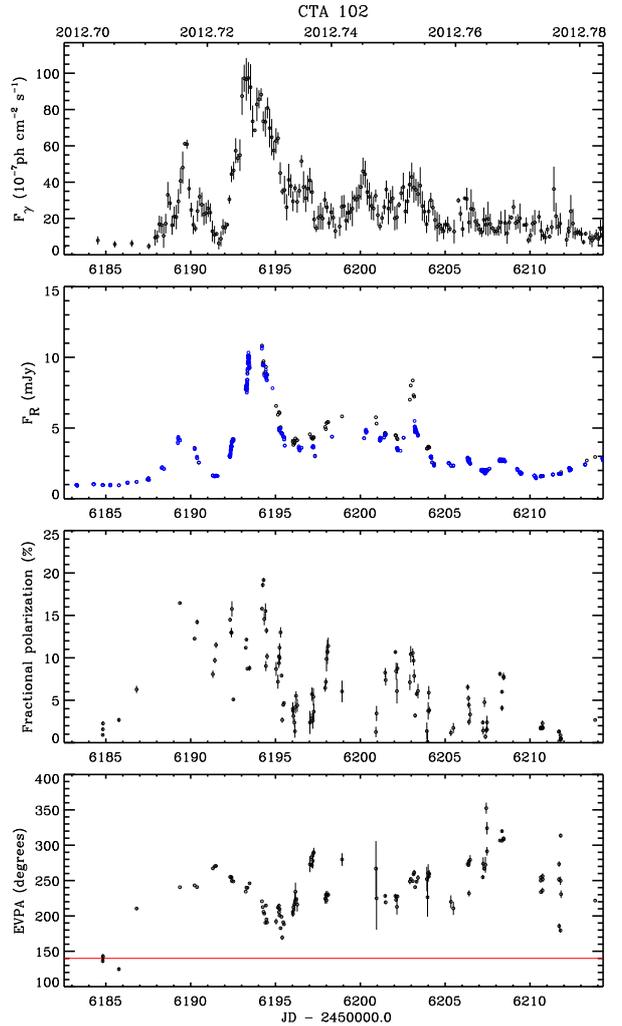}
        \caption{Blow-up of Fig.~\ref{cta102lc1} during the 2012 September-October flare.}
\label{lc_flare}
\end{center} 
\end{figure}

We use polarimetric data collected at St.~Petersburg University (Crimea and St.~Petersburg), Lowell (Perkins),
Steward, and Calar Alto observatories, supplementing these with data from the Kanata telescope \citep{2013ApJ...768L..24I}.
Instrumental polarization was derived from measurements of stars located near the object under the assumption that their
radiation is intrinsically
unpolarized. The Galactic latitude of CTA102 is -38\degr and $A_V$ = 0.16 mag, so that interstellar polarization (ISP) in
its direction is less than 0.6 per cent. To correct for ISP, the mean relative Stokes parameters of nearby stars were
subtracted from the relative Stokes parameters of the object. This removes the instrumental polarization as well.
The fractional polarization has been corrected for statistical bias, according to \citet{Wardle1974}.
Figure~\ref{cta102lc1} presents the flux and polarization behaviour of CTA~102 for 2005--2015. We supplement this plot
with a panel showing the $\gamma$-ray light curve from the \textit{Fermi} LAT in order to demonstrate that the most
prominent $\gamma$-ray activity ever recorded for this source was observed during the September--October 2012 optical
outburst. In Fig.~\ref{lc_flare} we present a blowup of the most active interval of the 2012 outburst. From visual
inspection of these figures, it is apparent that during the entire time range covered by \textit{Fermi}
observations up to the 2012 season, CTA~102 remained inactive at both $\gamma$-ray and optical bands; the degree of
polarization was mostly $\le10$ per cent, while the electric-vector position angle (EVPA) showed marked variations over the
range [-200\degr, 400\degr].
We resolve the $\pm180\degr$ ambiguity by adding/subtracting $180\degr$ each time that the subsequent value of the EVPA
is $>90\degr$ less/more than the preceding one. Occasional clockwise rotations of the polarization vector by up to
$\sim700\degr$ are apparent. The onset of the activity in the 2012 season was accompanied by a violent increase of optical
polarisation activity.  The degree of polarisation exceeded 20 per cent at some epochs, during which the position angle varied
over the range 150--$300\degr$.

\subsection{$\gamma$-ray Observations}
The $\gamma$-ray data were obtained with the {\em Fermi} (LAT),
which observes the entire sky every 3 hours at energies of 20 MeV--300 GeV \citep{Atwood2009}. We analysed the LAT data in
the energy range 0.1--200~GeV using the unbinned likelihood analysis of the standard
{\em Fermi} analysis software package Science Tools v9r33p0 and instrument response function {\tt P8R2\_SOURCE\_V6}. Source
class photons (\texttt{evclass}=128 and \texttt{evtype}=3) were selected within a $15\degr$ region of interest centred on
the blazar. Cuts in the satellite
zenith angle ($< 100\degr$) and rocking angle ($< 52\degr$) were used to exclude the Earth limb background. The
diffuse emission from the Galaxy was modelled using spatial model \texttt{gll\_iem\_v06}. The extragalactic diffuse and
residual instrumental backgrounds were included in the fit as an isotropic spectral template \texttt{iso\_source\_v05}. The
background model\footnote{\url{http://fermi.gsfc.nasa.gov/ssc/data/access/lat/4yr_catalog/gll_psc_v16.xml}}
includes all sources from the 3FGL catalog within $15\degr$ of the blazar.
Photon fluxes of sources beyond $10\degr$ from the blazar and spectral shapes of all targets were fixed to their values
reported in the 3FGL catalogue. The source is considered to be detected
if the test statistic TS provided by the analysis exceeds 10, which corresponds to approximately a $3\sigma$ detection
level \citep{Nolan2012}. The systematic uncertainties in the effective LAT area do not exceed 10 per cent
in the energy range we use \citep{Ackermann2012}. This makes them insignificant with respect to the statistical
errors, which dominate over the short time scales analysed in this paper. Moreover, our analysis is based on the relative
flux variations. Because of this, the systematic uncertainties are not taken into account.

Different time bins $t_{\rm int}$, from 12 hours to 7 days, were used, depending on the flux density of the object.
This maximizes the availability of detections at temporal resolutions that are as short as possible.

\subsection{\textit{Swift} Observations}
\subsubsection {Optical and Ultraviolet data}
{\em Swift} UVOT observations were performed in the optical $v$,
$b$, and $u$ bands, as well as in the UV filters $uvw1$, $uvm2$, and $uvw2$. 
We reduced the data with HEAsoft package version 6.10, with the 20101130 release of
the \textit{Swift}/UVOTA CALDB. Multiple exposures in the same filter
at the same epoch were summed with \texttt{uvotimsum}, and then
aperture photometry was performed with the task \texttt{uvotsource}. We used an aperture radius of 5\arcsec
centred on the source, and background from an annulus between 25\arcsec and 35\arcsec radii.
To take the spectral shape of CTA~102 into account, we re-calibrated the effective wavelengths and count-to-flux
conversion factors as explained in \citet{Raiteri2010}, using a power-law fit to the average source spectrum.
This also produced a better agreement between the ground-based and space data than when using the
\citet{Breeveld2011} calibrations. Galactic extinction was calculated by convolving the \citet{Cardelli1989} mean
extinction laws with the filter effective areas and source flux. All of the derived parameters are given in
Table~\ref{swift_calib}.

\begin{table}
\caption{\bf Swift calibrations used for  CTA~102 analysis.}
\label{swift_calib}
\begin{tabular}{c | c | c | c | c | c | c |}
\hline

 Bandpass & v & b &  u & uw1 & um2 & uw2 \\
\hline

$\lambda$, \AA & 5427 & 4353 & 3470 & 2595 & 2250 & 2066 \\
\hline
$A_\lambda$, mag & 0.24 & 0.32& 0.38 & 0.54 & 0.67 & 0.64 \\
\hline
conv. factors  &2.603 & 1.468 & 1.649 & 4.420 & 8.372 & 5.997 \\
\hline
\end{tabular}\\
\flushleft{Note -- Units of count rate to flux conversion factors are $10^{-16}{\rm erg}\,{\rm cm^{-2}s^{-1}}$\AA$^{-1}$.}

\end{table}

\subsubsection{X-ray Data}
The X-ray data were obtained over a photon energy range of 0.3$-$10 keV by the {\em Swift} XRT.
We reduced the data using HEAsoft package version 6.11. The standard \texttt{xrtpipeline} task was used to calibrate
and clean the events. We selected events with grades 0$-$12 in \texttt{pc} mode and 0$-$2 in \texttt{wt} mode.
An ancillary response file was created with a PSF correction using the \texttt{xrtmkarf} task, and the data
were rebinned with the \texttt{grppha} task to ensure a minimum of 10 photons in every newly defined channel. 
We fit the spectra with the spectral analysis tool \texttt{xspec} using a power-law model with minimum $\chi^2$
value and fixing the hydrogen column density ($N_{H} = 5.04\times10^{20} ~{\rm cm^{-2}}$) according to the
measurements of \cite{Dickey1990}. We used Cash statistics along with Monte Carlo spectral simulations to estimate
the goodness of fit at a confidence level of 90 per cent. If the parameters failed a goodness of fit test, we rebinned
the data with a minimum of 20 photons in each spectral channel and repeated the model-fit procedure. If the new model
still did not satisfy a goodness of the fit test, we rejected the data; this occurred only in 2 cases.

\section{Results}\label{results}
\subsection{Colour Evolution}\label{color_evolution}
The question of whether a blazar's radiation becomes redder or bluer when it brightens is a topic of numerous papers.
It is commonly agreed that the relative contributions of the big blue bump (BBB) and Doppler-boosted synchrotron
radiation from the jet are different between quiescence and outbursts,
and that this leads to variability of the spectral energy distribution (SED). The situation is even more complicated
in cases like CTA~102, where broad emission lines contaminate the wide photometric bands
(e.g., the \ion{Mg}{ii} $\lambda 2800$\AA\, line is redshifted to $\lambda 5700$\AA). A straightforward way
to isolate the contribution of the component of radiation that is variable on the shortest time-scales (presumably,
synchrotron radiation) has been suggested  by Hagen-Thorn \citep[see, e.g.,][and references therein]{Hagen-Thorn2008}.
 The method is based on plots of (quasi)simultaneous flux densities
in different colour bands and the construction of the relative continuum spectrum based on the slopes of the sets of flux-flux
relations thus obtained.

\begin{figure}
   \centering
  \includegraphics[width=\columnwidth,clip]{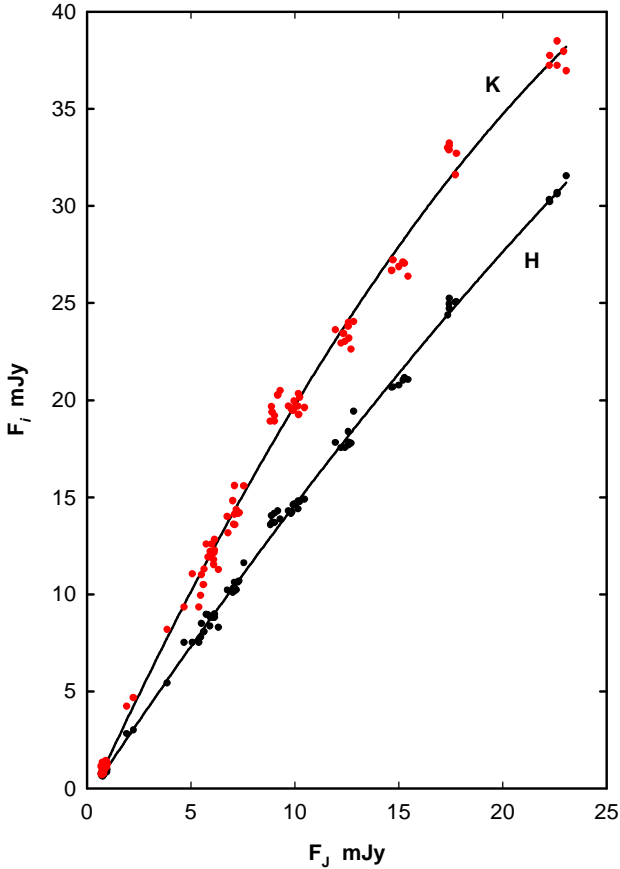}
      \caption{Flux-flux relations between the near-infrared $J$ band and the $H$ (black circles)  and $K$ (red circles) bands
      over the time interval 2008--2012. Lines are second order polynomial fits.}
         \label{NIR_ff}
   \end{figure}

\begin{figure}
   \centering
  \includegraphics[width=\columnwidth,clip]{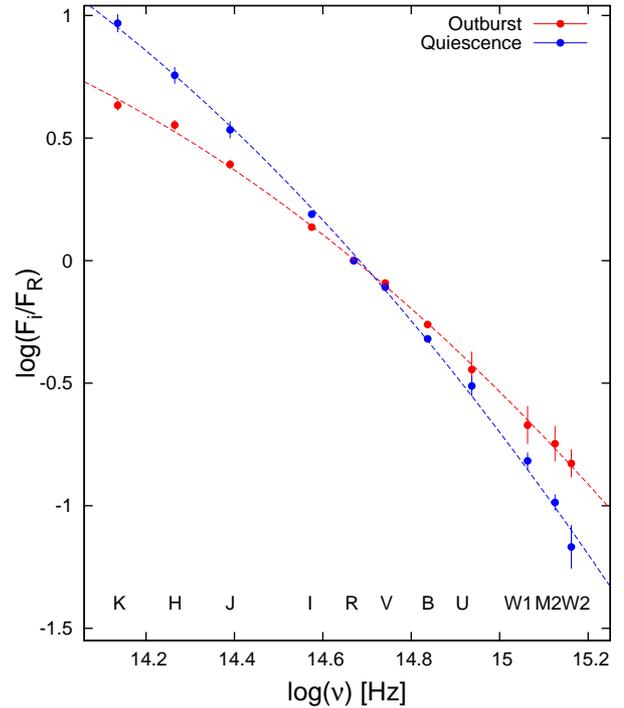}
      \caption{Relative continuum spectra of the \textit{variable} component in CTA~102 during quiescence (blue) and the 2012
      flare (red) from NIR to UV.}
         \label{SED_nir_uv}
   \end{figure}

An example of such an approach is given in Fig.~\ref{NIR_ff}, where the flux densities in $H$ and $K$ bands are
plotted against the $J$-band flux density. The lack of linearity between variations in corresponding bands means that
the low- and high-flux behaviours could reflect variability of different sources of radiation (e.g., the ambient
jet in low states and a shock in high states). Alternatively, if the same component is responsible for all of the 
variability patterns, the parameters of this component change significantly in a manner that depends
on the brightness of the source.
In Fig.~\ref{SED_nir_uv} we plot relative SEDs of the variable component in CTA~102 in quiescence and during the
2012 outburst from \textit{Swift} UV to NIR bands, showing marked hardening of the SED during the high state,
together with substantial curvature (convexity) of the spectrum. Since the logarithmic spectral shapes are far from
linear, we are not able to determine a single power-law slope $\alpha$ (in the sense $F_\nu \propto \nu^{-\alpha}$)
for the entire optical--NIR range. As a value that characterises these slopes, we select the tangent to the
spectrum at the central $R$-band frequency. For the quiescent stage, we obtain $\alpha_R=1.78\pm 0.05$, and for
the outburst $\alpha_R=1.50\pm 0.03$. We emphasize that these values refer to the \textit{variable} component only,
not to the entire flux. Meanwhile, if we look at the evolution of the \textit{total}-flux optical SED, we see the
opposite: $\alpha_R=0.4\pm 0.1$ during quiescence and $\alpha_R=1.4\pm 0.1$ for the outburst. The closeness of
the latter value to that obtained for the variable source is caused by the fact that, during the outburst state, the
relative contribution of intrinsically blue underlying components (BBB+ \ion{Mg}{ii} line emission) becomes
small compared to the synchrotron radiation of the variable source.

\begin{figure}
   \centering
  \includegraphics[width=\columnwidth,clip]{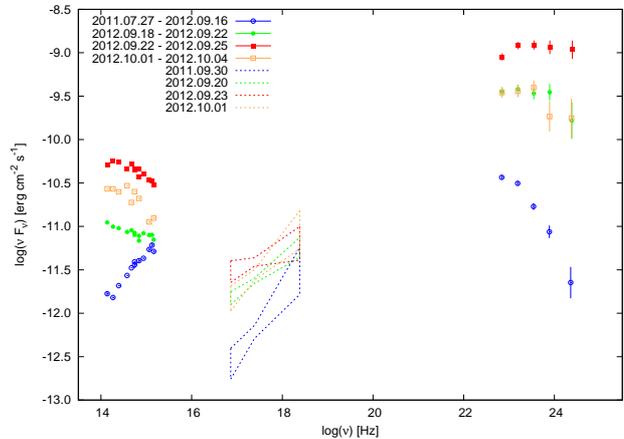}
      \caption{(Quasi)simultaneous SEDs of CTA~102 from NIR to $\gamma$-ray bands.}
         \label{SED_all}
   \end{figure}

Simultaneous spectral hardening in the $\gamma$-ray region during the outburst is also apparent in Fig.~\ref{SED_all}.
Notice that in this figure we plot total flux densities, in contrast to Fig.~\ref{SED_nir_uv}.

We hypothesize, as suggested in \citet{Larionov2010} for the case of BL~Lac, that this spectral hardening of the
variable optical and $\gamma$-ray components is
mostly caused by a change of the viewing angle of the emitting zone, which shifts in frequency the synchrotron
spectrum due to increased Doppler boosting. Some (or all) of the hardening could also result from the population of
emitting electrons becoming enriched with a high-energy extension during the outburst compared to the quiescent state.

\subsection{$\gamma$-ray -- Optical Correlations}
\label{gam_opt}

\begin{figure}
   \centering
  \includegraphics[width=\columnwidth,clip]{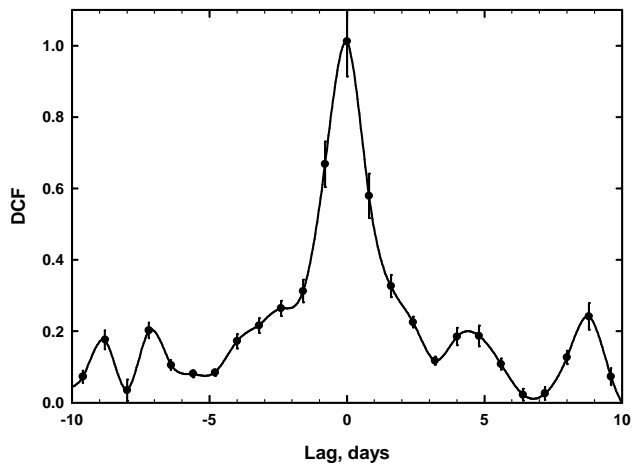}
      \caption{DCF between optical and $\gamma$-ray light curves of CTA~102. Negative delays correspond to
      $\gamma$-ray lagging behind optical variations. The zero delay at the peak of the DCF indicates co-spatiality of
      the active regions.}
         \label{dcf}
   \end{figure}

We have calculated the discrete correlation function (DCF) \citep{1988ApJ...333..646E, Hufnagel1992} between the
optical and $\gamma$-ray flux variations of CTA~102 during 2012. The results, given in Fig.~\ref{dcf},
clearly demonstrate that there is no time delay between the variations in the two energy bands within the
accuracy of the DCF method. The value of the lag between optical and $\gamma$-ray variations, based on the DCF
centroid position, is $-0\fd05\pm0\fd02$. One may note that there are secondary `humps' at $\approx 4.5$ days and
$\approx 9$ days. We surmise that these are caused by recurring optical and $\gamma$-ray sub-flares during
the 2012 outburst (see \S~\ref{pol_behaviour} and Fig.~\ref{model} below).

\begin{figure}
\begin{center}
   \includegraphics[width=\columnwidth,clip]{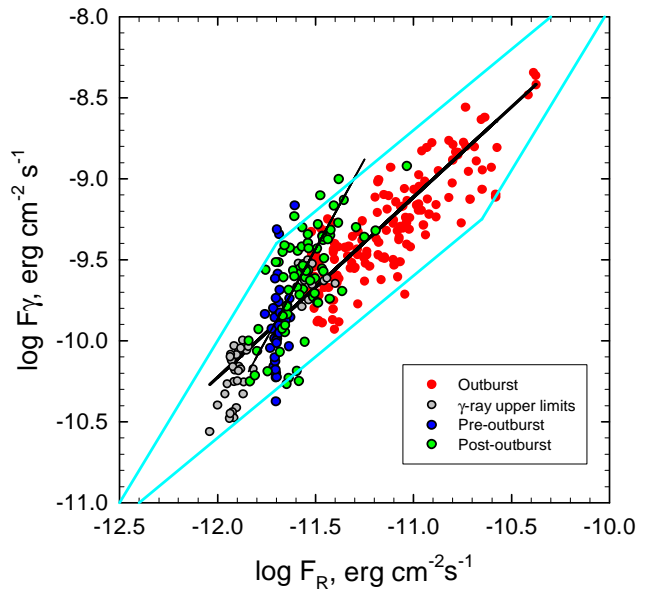}
        \caption{$\gamma$-ray -- optical flux-flux diagram. Slopes of the linear (on a logarithmic scale) fits
        are $1.12\pm0.04$ (outburst stage) and $2.21\pm 0.32$ (post-outburst stage). Almost all data points lie inside
        the parallelogram, whose sides have slopes of 1 and 2 (see text).}
\label{gamma-optical}
\end{center} 
\end{figure}

This lack of delay allows us to compare directly the optical and $\gamma$-ray flux densities. To do this, we
(1) bin the $R$-band optical data so that the mid-point and size of each optical bin corresponds to the mid-point and
size of the respective $\gamma$-ray bin, and (2) subtract from the binned optical data a tentative value of
the flux of  (quasi-)permanent emission components (BBB + QSO-like emission with a prominent \ion{Mg}{ii} line).
Combined, this amounts to $\log(\nu F_\nu)=-11.5$ in $R$ band, which is similar to the value obtained for CTA~102
in \citet{Raiteri2014}, corresponding to as much as 50 percent of the total quiescent flux. Figure~\ref{gamma-optical}
demonstrates clear differences during the various stages of CTA~102 activity. The onset of $\gamma$-ray activity
(TJD~5700--5943, blue circles in Fig.~\ref{gamma-optical}) corresponds to a rather stable optical level. During
the outburst stage (TJD~6069--6678, red circles), we see a relation between  $\gamma$-ray and optical fluxes of the form
$F_\gamma \propto F_{\mathrm opt}^{1.12\pm0.04}$, while in the post-outburst stage, TJD~ 6776--7231 (green circles),
$F_\gamma \propto F_{\mathrm opt}^{2.21\pm0.32}$.

We assume that the variable optical emission is mostly synchrotron radiation from 
the jet, while the $\gamma$-ray emission is from inverse Compton scattering (IC)
of optical/IR photons by the jet's relativistic electrons. The origin of the seed photons may
be external to the jet, e.g., hot dust continuum or broad line emission (external
Compton, or EC model), or synchrotron photons from the
jet (synchrotron self-Compton, or SSC model). In the EC model we expect  
the respective fluxes to vary as $F_{C} \propto F_{sync}$, since only 
the relativistic electron population is in common, while in the SSC model
$F_{C} \propto F_{sync}^{2}$, since both the relativistic electrons and emission radiated 
by them are involved in the high-energy photon production. Here, $F_{sync}$ is the flux 
of the synchrotron radiation and $F_{C}$ is that of the IC emission. These dependences will be altered
slightly by the different optical and $\gamma$-ray K corrections at times when the optical and $\gamma$-ray
spectral indices are not the same. 

A competing explanation of the near-unity slope between optical and $\gamma$-ray fluxes, besides the EC model,
is that their variability is mostly caused by variations of Doppler factor resulting from changes in the viewing angle.
This can occur if the entire jet changes its direction (wobbles or precesses), or if different parts of the jet
cross-section with various velocity vectors relative to the mean become periodically or sporadically bright as
time passes. Under this scenario,
the post-outburst stage with presumably small variations in viewing angle produces SSC-like variability
that was hidden during the height of the outburst under higher-amplitude Doppler-boosted variations of geometrical
origin.  

Thus, the data distribution in Fig.~\ref{gamma-optical} can be explained as due to two concurrent effects, with slopes
of $\sim 1$ (Doppler factor variations) and $\sim 2$ (intrinsic SSC dependence), so that the data mostly lie inside
a circumscribed parallelogram with sides having the above slopes. The relative lengths of these sides depend on the
relative dominance of the two effects, and the best-fit slope of the entire distribution can vary from $\sim 1$ to 2.
The best fit slope of 1.12 during the outburst would indicate almost complete dominance of the Doppler factor variations,
while the best fit found for the post-outburst stage implies an essentially constant, enhanced Doppler
factor during that period.
Another advantage of this model is that it can explain the polarimetric variability in
CTA~102 (see \S~\ref{pol_behaviour}).

\subsection{Optical Spectra}
\label{spectra}

\begin{figure}
   \centering
  \includegraphics[width=\columnwidth,clip]{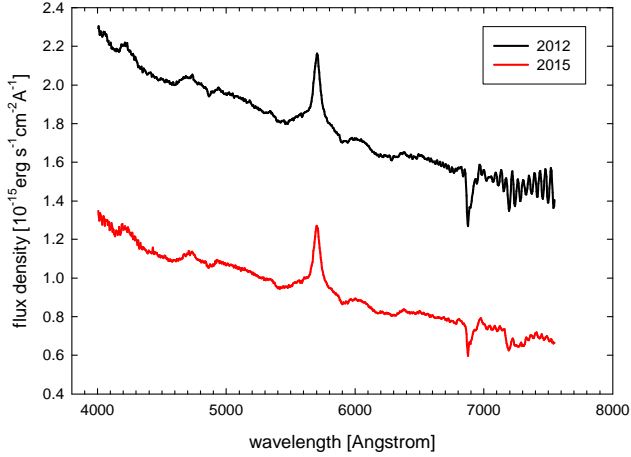}
      \caption{Averaged spectra of CTA~102 during the 2012 and 2015 observing seasons. Absorption features longward
       of $\lambda$6800\AA ~are of telluric origin.}
         \label{mean_spectra}
   \end{figure}

We analyse the optical spectroscopic behaviour of CTA~102 using the data taken at Steward Observatory
of the University of Arizona for the `Ground-based Observational Support of the {\it Fermi} Gamma-ray Space Telescope'
program\footnote{\url{http://james.as.arizona.edu/~psmith/Fermi/}} at the 2.3~m Bok telescope and 1.54~m Kuiper telescope
from 2008 to 2015. All of these spectra contain a prominent \ion{Mg}{ii} $\lambda 2800$\AA\, broad emission line redshifted
to $\lambda 5700$\AA. Figure~\ref{mean_spectra} displays averaged spectra for the 2012 and 2015 observing seasons.
We use 133 spectra spread over the time interval 2009--2015 to check whether there is any correlation between continuum
(mostly synchrotron) flux density variations and changes in the \ion{Mg}{ii} line flux. 

\begin{figure}
   \centering
  \includegraphics[width=\columnwidth,clip]{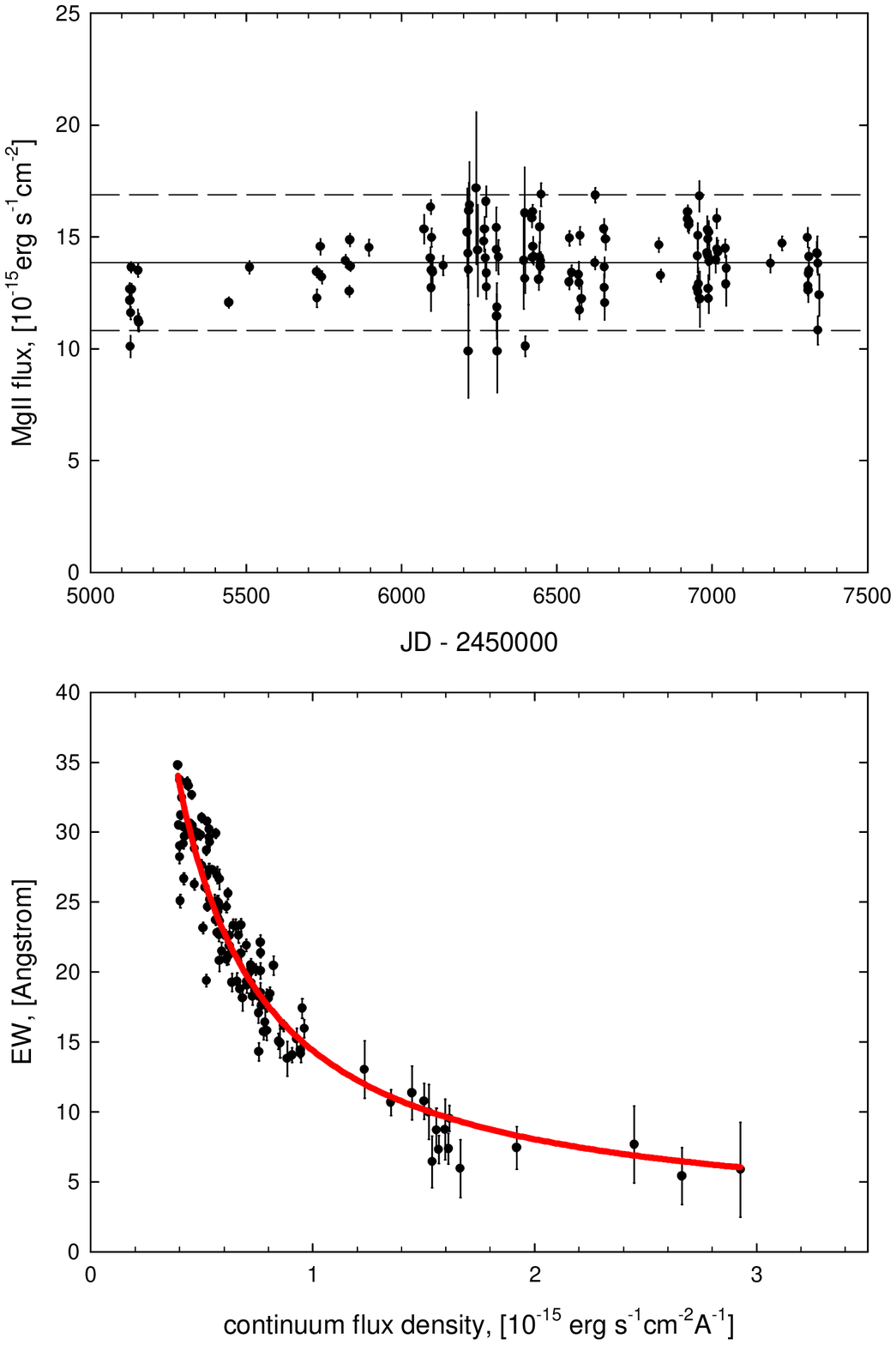}
      \caption{Equivalent width of
      \ion{Mg}{ii} line vs.\ continuum flux over the time interval 2009--2015.  The red curve corresponds to the expected EW if the line flux remains constant.}
         \label{flux_ew}
   \end{figure}

We evaluate the line parameters (the equivalent width, EW, and the line full width at half-maximum, FWHM), fitting the line
profile with a single Gaussian function superposed on a featureless continuum. 
The results are presented in Fig.~\ref{flux_ew}, where EW is plotted against the continuum flux density; the inverse proportionality of these
two quantities indicates that the line flux is stable. The red curve corresponds to the expected EW if the line flux remains constant. 
These results imply that enhanced activity of
the jet has little or no effect on the broad-line region (BLR), where one expects most or all of the \ion{Mg}{ii} emission to originate.

We note that similar results have been obtained for other blazars, e.g., 3C~454.3~\citep{Raiteri2008} and
OJ~248 \citep{Carnerero2015}. However, some cases of correlated broad-line flux variability connected to $\gamma$-ray
variability have indeed been reported by \citet{Leon-Tavares2013}, \citet{Isler2013}, and \citet{Isler2015}.

We measure the \ion{Mg}{ii} line FWHM, from which one can derive the velocity of the gas clouds in the BLR, and
obtain $v_\mathrm{FWHM} = 2100\pm250\:\mathrm{km\: s^{-1}}$. This value is a lower limit to the actual
velocity range of the broad-line clouds, since it depends on the geometry and orientation of the BLR~\citep[see, e.g.,][]{Wills1995}. In fact,
because the line of sight to a blazar is probably nearly perpendicular to the accretion disk, the de-projected
velocity range is likely to be a factor $\gtrsim 2$ higher than the FWHM given above.

\section{Discussion}\label{discussion}

\subsection{Polarimetric Behaviour and Helical Jet Model }\label{pol_behaviour}

\begin{figure}
   \centering
  \includegraphics[width=\columnwidth,clip]{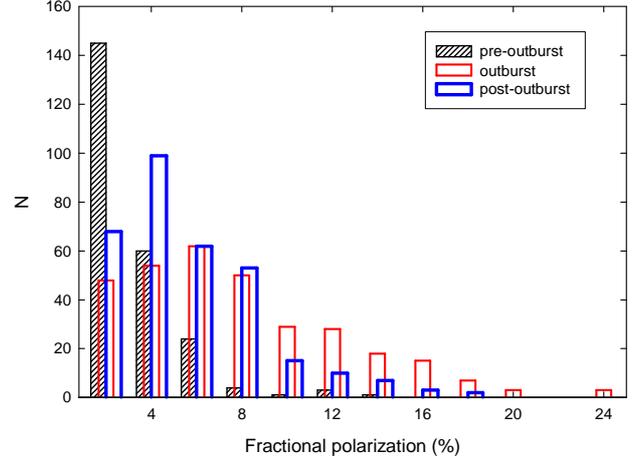}
      \caption{
			Histograms of fractional polarization before (black shaded), during (red), and after (blue)
      the 2012 outburst. 
			}
         \label{histo}
   \end{figure}

\begin{figure}
   \centering
  \includegraphics[width=\columnwidth,clip]{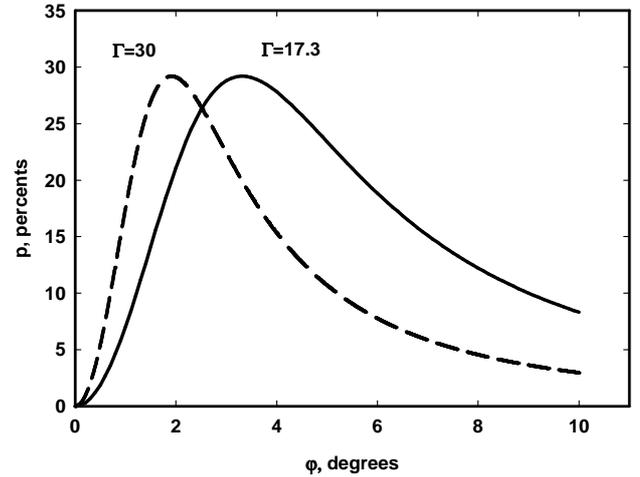}
      \caption{Behaviour of fractional polarization vs.\  viewing angle for plasma compression
      ratio $\eta=1.5$, Lorentz-factor $\Gamma = 17.3$ (solid line), and $\Gamma = 30$ (dashed line), in the moving shock model.}
         \label{pol_vangle}
   \end{figure}

Our polarimetric data obtained during 2008-2015 allow one to see remarkable changes in the behaviour of CTA~102
that were probably triggered by (or, at least, coincided with) the prominent outburst of 2012. Figure~\ref{histo}
shows histograms of the polarization degree (PD) before, during, and after the flare. This highlights
the increased range of PD variations already seen in Fig.~\ref{cta102lc1}. A natural reason for this change is
a decrease in the viewing angle of the jet, as already suggested by \citet{Casadio2015} based on analysis of the
superluminal apparent motion of knots in VLBA images. However, if we consider the range of values of viewing angles
found in that paper (from $3.9\degr$ before the 2012 outburst to $1.2\degr$ after it) and compare the values of PD
expected within the moving shock model for polarization variations
(see, e.g., Fig.~\ref{pol_vangle} and also \citet{Larionov2013, Raiteri2013}), we find that
we would expect to see the opposite: a decrease in PD during the
outburst. A positive correlation between the photometric flux and PD may be obtained if the bulk
Lorentz factor of the emitting plasma is much higher, e.g., $\Gamma\approx30$ (dashed line in Fig.~\ref{pol_vangle}). In this case, a decrease in viewing angle would increase PD \citep[see also Eq.~(1)-(3) in][]{Larionov2013}. However, such a high value is difficult to reconcile with that found by \citet{Casadio2015}, $\Gamma=17.3$. 

Yet another possible reason for this apparent contradiction could be 
the difference in sizes 
between the parts of jet
responsible for the flaring optical radiation and the centroid of the radio `core'.
In this case, the source of the polarized optical flux could have a mean
velocity vector that is less well aligned with the line of sight than that of the radio emission region. This explanation is supported by very different time scales of variability  in optical (few days) and radio (months) wavelengths and, correspondingly, different sizes of the emission regions \citep[see also][]{Casadio2015}.

\begin{figure}
   \centering
  \includegraphics[width=\columnwidth,clip]{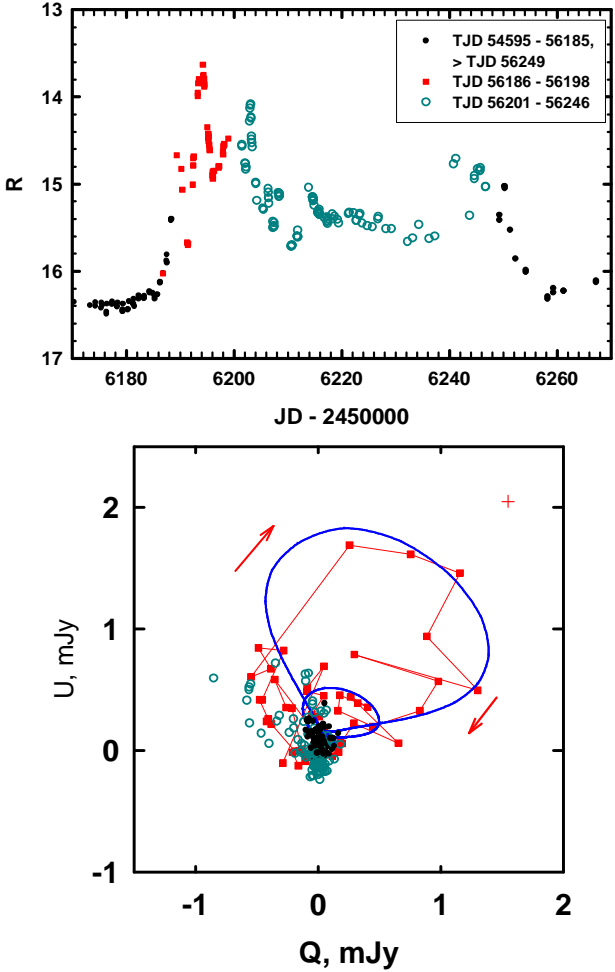}
      \caption{Bottom panel: absolute Stokes parameters $Q$ and $U$  of the optical polarization of CTA~102.
      The typical errors in the Stokes parameters are shown by the red cross. The blue curve corresponds to
      the path of the vector of polarization according to our model (see text). The top panel presents
      the $R$-band light curve, with intervals of time colour-coded in the
      same way as in the plot of Stokes parameters.}
         \label{qu}
   \end{figure}

Figure~\ref{qu} shows the distribution of the absolute Stokes parameters of CTA~102 during both quiescence and
different stages of the 2012 activity. We notice that the cluster of $(Q,U)$ points obtained
before and after the outburst (more than 300 data points, marked as black circles) is located near the origin of
the coordinates. All of the data points are tightly packed around this location, which corresponds to a very
low level of polarized flux during quiescence (see also Figs.~\ref{cta102lc1}~and~\ref{histo}). The onset of the outburst was
accompanied by a definite loop-like rotation in the plane of the Stokes parameters, while the fading stage of the
outburst included less ordered drifts, misplaced relative to the pre-outburst position. This latter feature may
reflect the change in orientation of the jet itself, while the clockwise rotation could arise from
spiral movement of the radiating blob through the jet. 

As in the case of S5~0716+71~\citep{Larionov2013}, we propose a model of a relativistic shock moving down a helical jet,
or along helical magnetic field lines, to explain these rotations. 
The main parameters that determine the visible  behaviour of the outburst are:
(1) jet viewing angle $\theta$; (2) pitch angle $\zeta$ of the spiral motion and helical field;
(3) parameters of polarization of the undisturbed jet; (4) bulk Lorentz factor $\Gamma$ of the shocked plasma;
(5) scaling factor of the exponential rise of the outburst $\tau$; (6) factor $k$, responsible for different
time-scales of the rise and decline of the outburst; (7) period of the shock's spiral revolution in the observer's
frame $P_\mathrm{obs}$; (8) the same period in the source frame; (9) radius of spiral; (10) shocked plasma compression
$\eta$ = ratio of post-shock to pre-shock density; and (11) spectral index of the emitting plasma $\alpha$. 

Some of these parameters can be obtained, or at least constrained, directly from observations. For example, the
Lorentz factor $\Gamma$ during the 2012 outburst is close to 17, according to \citet{Casadio2015};
the average level of polarization during quiescence is of order 1 per cent (see Fig.~\ref{histo}); the value of
$P_\mathrm{obs}\approx4\fd7$ is obtained from repeating optical (and $\gamma$-ray) sub-flares during the
outburst; the mean value of the slope of the synchrotron spectrum $\alpha=1.50$, which we obtain from our
photometric data (see \S~\ref{SED_nir_uv} and Fig.~\ref{SED_nir_uv}).

\begin{figure}
   \centering
  \includegraphics[width=\columnwidth,clip]{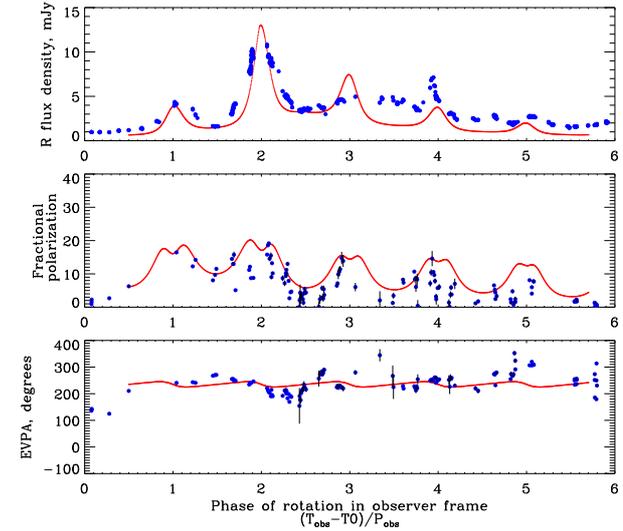}
      \caption{Comparison of optical photometric and polarimetric data during the giant outburst of CTA~102 in 2012,
      with our model fit.}
         \label{model}
   \end{figure}

Using relations (1)-(9) from \citet{Larionov2013}, we obtain the values of the model parameters that are given in
Table~\ref{tab_1}. To confront the model with observational results, we plot both together in Figs.~\ref{qu} and
\ref{model}.
Since our model only takes into account smooth variability caused by a radiating blob moving along a helical path
in the jet and neglects the effects caused by turbulence that is probably present, we are able to reproduce only
the basic variability pattern. In particular, we see a series of decaying flares after the main outburst (and the
precursor preceding it). Nevertheless, the agreement of the model with the $Q$ vs.\ $U$ evolution in Fig.~\ref{qu}
is quite good. The model evolution of the degree of polarization corresponds to an upper envelope to
the observational data. This is as expected, since turbulence adds superposed polarization vectors at random
position angles, which often partially cancel the polarization from the ordered component of the magnetic field.

\begin{table}
\tiny
\caption{\bf Model parameters for the photometric and polarimetric behaviour of CTA~102 in 2012 September.}
\label{tab_1}
\begin{tabular}{ccccccccccc}
\hline

$\theta\degr$ & $\zeta\degr$ & $p_\mathrm{jet}$ & $\Gamma$ &  $\tau$ & $k$ & $P_\mathrm{obs}$ & $P_\mathrm{src}$ & $r$  &  $\eta$ & $\alpha$\\
 (1) & (2) & (3) & (4) & (5) & (6) & (7) & (8) & (9) & (10) & (11) \\
\hline

2.65 & 0.9 & 1 & 18.2 & 0.75 & 1.83 & 4.7 & 2.35 & 0.0018 & 1.35 & 1.50 \\
\hline
\end{tabular}\\
\flushleft{Note. -- Units: $p_\mathrm{jet}$ in per cent, $r$ in parsecs, $P_\mathrm{obs}$ in days, $P_\mathrm{src}$ in years. $\tau$ and $ k$ in
fractions of $P_\mathrm{src}$.}

\end{table}

We note that our finding of clockwise rotation of the polarization vector is supported by the detection of negative
circular polarization in the 15~GHz radio emission of CTA~102 by \citet{2008MNRAS.384.1003G}, who used the
observation as evidence for a helical magnetic field. In addition, inspection of the EVPA behaviour in
Fig.~\ref{cta102lc1} allows one to see at least 3 episodes of clockwise rotation with amplitude exceeding
700\degr (TJD~5500, 6500, 7250) and no cases of counter-clockwise rotations of similar length. 
Thus, this appears to be a persistent feature of the blazar, in agreement with an ordered, helical component of the magnetic field.

\subsection{Weakness of Spectral Variations}

The \ion{Mg}{ii} emission line flux is, at most, weakly variable over the course of our observations despite marked
changes in the optical synchrotron flux. We consider this to be a consequence of two factors: (1) the part of jet
where the outburst occurred, is located parsecs outside of the BLR \citep[see][]{Casadio2015}, and (2) the ultra-high
amplitude of the 2012 outburst might have been determined to major extent by a change in the viewing angle.
The number of ionizing photons traversing the BLR would not significantly change during such a re-orientation
of the jet.

\subsection{Implication of Variable Doppler Factor}
We have used our photometric and polarimetric data centred on the 2012 outburst to assess the main geometric parameters
that govern the overall pattern of Stokes parameter variations within a model of a radiating blob or shock wave that moves
along a helical path down the jet. The changes in the viewing angle caused by helical motion imply strong changes in
the Doppler factor, from $\delta\approx 28$, when the emission region is closest to our line of sight,
to $\delta\approx 16$ when it is farthest. The larger value is similar to $\delta_{var}\approx30$ obtained
by \citet{Casadio2015} based on VLBA data.

 Our long-term study of the polarimetric behaviour of CTA~102 allows us to identify at
least 3 episodes of sustained clockwise rotations and no similar episodes of counter-clockwise rotation. This repeated
behaviour suggests that the cause of the rotations is geometrical rather than the result of random walks related to a disordered
magnetic field (although some shorter, apparently random rotations, occur as well)~\citep{Kiehlmann2016}.

As is shown in Figs.~\ref{cta102lc1},~\ref{histo}, the mean level of polarization of CTA~102 substantially increased
shortly before the onset of the 2012 outburst, and did not revert to the pre-outburst level until fall 2015. This
supports the hypothesis that secular variations of the viewing angle of the jet led to both enhanced photometric
activity and corresponding changes in the linear polarization.
We can expect that flaring activity in this blazar will be more pronounced
than in previous years as long as the jet remains closely aligned with our line of sight. Indeed, as reported by \citet{Carrasco2016}, \citet{Balonek2016}, and \citet{Becerra2016},
at the end of 2016 January a new high-amplitude outburst occurred at $\gamma$-ray, optical, and near-infrared bands (see
also \url{http://vo.astro.spbu.ru/sites/default/files/optic/cta102R.png}). Unfortunately, this happened when the object was
difficult to observe owing to proximity to the sun in the sky, so this expected effect cannot yet be verified in detail.

\section{Conclusions}\label{conclusions}

During the GASP/WEBT campaign we obtained densely sampled optical photometric and polarimetric data around the
period of unprecedented optical and $\gamma$-ray activity of CTA~102, and combined optical data with
contemporaneous observations throughout the $\gamma$-ray to near-infrared frequency range. 
We find detailed correspondence of optical and $\gamma$-ray events, which confirms co-spatiality of the synchrotron
and inverse Compton emission sites. The relation between optical and $\gamma$-ray flux during the height of the outburst
is roughly linear. This is as expected from either the external Compton process for the high-energy emission or from
variable Doppler boosting acting as the main factor controlling the overall pattern of variability at both energy ranges. However, the Doppler boosting caused by changed viewing angle of the emission region is a preferred explanation for the variability of the total flux and polarization parameters.
In contrast, during the decay the relation between the fluxes is, within the uncertainties, consistent with the
$F_\gamma \propto F^2_{\mathrm{opt}}$ law expected from the SSC mechanism. Presumably, any changes in viewing angle
during the decline were too minor to have a dominant effect on the variations in flux. 

We have determined the SED of the variable component of synchrotron emission during both quiescence and the stages
of outburst, and found appreciable hardening of the SED during the outburst. This hardening could be explained by
convexity of the UV--NIR spectrum (see Fig.~\ref{SED_nir_uv}) that moved to higher frequencies owing to an increased
Doppler shift as the viewing angle decreased. This effect could have been amplified by an increase in the number
of high-energy electrons. The same spectral hardening is apparent in the $\gamma$-ray part of the spectrum.

As we can judge from our data, the change of viewing angle that led to enhanced activity in CTA~102 starting in 2012
may have resulted in a higher duty cycle of activity. When the viewing angle is smaller, the Doppler factor is more
sensitive to changes in that angle, hence variations caused by a non-constant viewing angle will be more pronounced
and occur over shorter time intervals. 

\section*{Acknowledgements}
The St. Petersburg University team acknowledges support from Russian RFBR 
grant 15-02-00949 and St. Petersburg University research 
grants 6.38.335.2015, 6.42.1113.2016. The research at Boston University was funded in part by NASA Fermi Guest Investigator grants NNX08AV65G, NNX10AO59G, NNX10AU15G, NNX11AO37G,
NNX11AQ03G, and NNX14AQ58G. The research at Steward Observatory was funded in part by NASA Fermi Guest Investigator grants NNX09AU10G and NNX12AO93G. Acquisition of the MAPCAT data at Calar Alto is performed at the IAA-CSIC and is supported by the Spanish Ministry of Economy and Competitiveness (MINECO) grant AYA2013- 40825-P. I.A. acknowledges support by a Ram\'on y Cajal grant of the MINECO. AZT-24 observations are made within an agreement between  Pulkovo, Rome and Teramo observatories. The PRISM camera at Lowell Observatory was developed by K. Janes et al. at BU and Lowell Observatory, with funding from the NSF, BU, and Lowell Observatory. This research was conducted in part using the Mimir instrument, jointly developed at Boston University and Lowell Observatory and supported by NASA, NSF, and the W. M. Keck Foundation. Calar Alto Observatory is jointly operated by the Max-Planck-Institut f\"ur Astronomie and the Instituto de Astrof\'isica de Andaluc\'ia-CSIC. This research was partially supported by the Scientific Research Fund of the
Bulgarian Ministry of Education and Sciences under grant DO 02-137
(BIn-13/09).  The Maidanak Observatory team acknowledges support from Uzbekistan Academy of Sciences grant F2-FA-F027. Skinakas Observatory is a collaborative project of the
University of Crete, the Foundation for Research and Technology -- Hellas,
and the Max-Planck-Institut f\"ur Extraterrestrische Physik. This article is partly based on observations made with the IAC80
and TCS telescopes operated by the Instituto de Astrofisica de Canarias in the Spanish
Observatorio del Teide on the island of Tenerife.

\bibliographystyle{mnras} 
\bibliography{cta102}

%
%

\bsp
\label{lastpage}
\end{document}